\documentclass[cameraready]{Interspeech}
\newcommand{\newpara}[1]{\vspace{5pt}\noindent\textbf{#1}}
\usepackage{multirow}
\usepackage{url,cite}
\usepackage{booktabs}
\usepackage{makecell}
\usepackage{enumitem}
\usepackage{amsmath}
\title{Who Spoke When in Multi-Conversation:\\ Target Speaker Tagging Task and Benchmark}

\author[affiliation={1}]{Minjae}{Lee}
\author[affiliation={1}]{Hee-Soo}{Heo}
\author[affiliation={1}]{Youngki}{Kwon}
\author[affiliation={1}]{Han-Gyu}{Kim}
\author[affiliation={2}]{You Jin}{Kim}
\author[affiliation={1}]{Bong-Jin}{Lee}

\address{
    $^1$ NAVER Cloud Corporation, South Korea
    $^2$ NAVER Corporation, South Korea
}

\email{mjlee.0328@navercorp.com}


\usepackage{comment}

\begin{document}

\maketitle

\begin{abstract}
We present target speaker tagging (TST), a task that integrates speaker diarization, verification, and identification into a unified workflow for multi-speaker conversations. Given long recordings and pre-enrolled speakers, TST detects and labels speech segments of known speakers while rejecting unknown ones. Despite its practical importance, research has been limited by the absence of suitable evaluation resources. To address this, we introduce TST-Bench, a large-scale synthetic benchmark with over 150 enrolled speakers, 300 sessions of 20--60 minutes, and reference annotations with global speaker labels. We define an evaluation protocol encompassing diarization and full-pipeline scenarios. Experiments on both real and synthetic data show that TST poses challenges not captured by conventional benchmarks, and that dedicated system design yields significant gains over naive integration of existing solutions. The benchmark dataset and evaluation protocols are publicly released.
\end{abstract}

%==================================================================================
\section{Introduction}
\label{sec:intro}

Speaker recognition is a broad research field encompassing tasks such as speaker verification, speaker identification, and speaker diarization~\cite{campbell2002speaker, bai2021speaker, dehak2010front, variani2014deep, park2022review, fujita2019end}, each addressing distinct aspects of the problem. Speaker verification determines whether two utterances originate from the same person. Speaker identification classifies a test utterance as one of the known speakers in a predefined set. Speaker diarization segments an audio recording by speaker without assigning identities. Although these tasks share core technologies such as speaker embeddings~\cite{snyder2018x, rouvier2015speaker, wang2021dku, park2019auto, anguera2012speaker, desplanques2020ecapa}, they have largely been studied in isolation.

While state-of-the-art speaker embedding extractors achieve remarkably low equal error rates on standard verification benchmarks such as VoxCeleb~\cite{Nagrani17, chung2018voxceleb2, Nagrani19}, evaluation based solely on speaker verification metrics captures only a narrow aspect of speaker recognition capability. In practice, real-world applications such as meeting transcription, voice-based services, and multi-session analytics require systems that simultaneously segment audio, identify speakers from a potentially large set of enrolled individuals, and reject unknown speakers---challenges that isolated benchmarks cannot adequately assess.

To bridge this gap, we propose \textit{target speaker tagging} (TST), a task designed to capture the full complexity of real-world speaker recognition. In TST, long audio recordings containing multiple speakers are given along with a set of pre-enrolled target speakers. The objective is to (1)~segment the audio into single-speaker regions, (2)~assign globally consistent speaker identities to segments belonging to enrolled speakers, and (3)~tag the remaining segments as non-target. Formally:
\begin{enumerate}
\item We assume long audio recordings in which multiple speakers may appear, as in speaker diarization.
\item A subset of speakers have enrolled their voice in the system in advance, as in speaker verification.
\item The objective is to detect and tag segments where enrolled speakers speak and assign the corresponding speaker identity, while labeling others as non-target.
\end{enumerate}

TST is not merely a concatenation of existing tasks but a complex problem with unique challenges. Certain diarization errors, such as merging distinct speakers into the same cluster, severely impact downstream identification. Conversely, over-clustering errors can sometimes be corrected at the identification stage. These interactions mean that simply assembling off-the-shelf systems does not yield optimal results; instead, each component must be adapted to the requirements of the integrated pipeline.

A critical barrier to advancing TST research is the \textit{absence of suitable evaluation resources}. Existing speaker diarization corpora present several limitations for TST evaluation:
\begin{itemize}
\item \textbf{Lack of global speaker labels.} Most diarization datasets assign session-local labels (e.g., Speaker~0, Speaker~1), making cross-session speaker identification impossible.
\item \textbf{Limited speaker populations.} Even datasets with global identifiers typically contain too few speakers to study the scalability challenges that arise when the enrolled population grows.
\item \textbf{No established evaluation protocol.} There is no standardized methodology for jointly evaluating diarization quality, identification accuracy, and unknown speaker rejection in an integrated framework.
\end{itemize}

This paper addresses these gaps comprehensively. Our contributions are:
\begin{enumerate}
\item We formalize target speaker tagging as a unified speaker recognition task that integrates diarization, identification, and verification, with a precise task definition and evaluation framework.
\item We propose a system specifically adapted for TST, demonstrating that dedicated design outperforms naive integration of existing methods.
\item We construct and publicly release \textbf{TST-Bench}, a large-scale synthetic benchmark featuring over 150 enrolled speakers, 300 multi-speaker sessions with configurable conditions, and reference annotations in RTTM format with global speaker labels.
\item We define an evaluation protocol that, unlike conventional speaker verification benchmarks relying on fixed pre-segmented utterances, accounts for variability in segmentation arising from different diarization system outputs.
\end{enumerate}

%==================================================================================
\section{Related Work}
\label{sec:related}

\newpara{Speaker diarization.} Speaker diarization has been extensively studied, from early HMM-based approaches~\cite{anguera2012speaker} to recent end-to-end neural methods~\cite{fujita2019end}. While modern systems achieve impressive DER on standard benchmarks, they produce only session-local anonymous labels, limiting their utility for applications requiring cross-session speaker identity.

\newpara{Speaker verification and identification.} Speaker verification determines whether two utterances share the same identity, with deep embedding approaches achieving sub-1\% EER on VoxCeleb benchmarks~\cite{Nagrani17, chung2018voxceleb2, Nagrani19}. Open-set speaker identification extends this to the multi-class setting, where the system must both select the correct speaker from a gallery and reject unknown speakers~\cite{malegaonkar2011performance}. However, these evaluations assume pre-segmented, clean utterances---a condition rarely met in practice.

\newpara{Integrated speaker recognition.} Despite the practical need, few studies have addressed the integration of diarization and speaker identification as a single pipeline. VoxBlink2~\cite{lin2024voxblink2} provides a large-scale evaluation for open-set identification but assumes pre-segmented utterances, leaving diarization-dependent scenarios unaddressed. To the best of our knowledge, no prior work has formally defined a unified task that jointly evaluates diarization, identification, and unknown speaker rejection, nor provided a dedicated benchmark for such evaluation. Our work fills this gap by formalizing TST and introducing TST-Bench.

%==================================================================================
\section{Target Speaker Tagging Framework}
\label{sec:framework}

In this section, we describe the system used for target speaker tagging. Although the main modules---speaker diarization and open-set speaker identification---are based on conventional speaker recognition techniques, we adapt and arrange them to meet the specific needs of TST.

\subsection{Overview}
\label{sec:overview}

The main objective of target speaker tagging is to convert the generic speaker labels produced by a diarization system (e.g., ``Participant~A'') into actual speaker identities drawn from a set of known speakers. Unlike standard speaker diarization, which operates within a single session, TST assigns globally meaningful identities across multiple sessions. It also differs from conventional speaker verification or identification, where clean single-speaker segments of sufficient length are assumed to be readily available.

It is important to distinguish between the \textit{TST system} and the \textit{TST scenario}. The TST system itself consists of two core modules---speaker diarization and open-set speaker identification---and takes as input a multi-speaker audio recording along with pre-enrolled speaker representations. Enrollment data, i.e., representative utterances of known target speakers, is provided externally and falls outside the system's computational pipeline. However, the complete TST scenario necessarily includes the enrollment step: without it, the system has no reference to identify speakers against. In a real-world deployment, a user would listen to a recorded session, identify segments containing their own speech, and submit them as enrollment utterances. Crucially, enrollment is a one-time process per speaker: once a speaker has enrolled through any session, the resulting representation is stored and reused across all subsequent sessions, eliminating the need for repeated enrollment. Additionally, the user specifies which enrolled speakers are expected to appear in each session, allowing the system to focus on a relevant subset of the gallery.

\begin{figure*}[t!]
    \centering
    \includegraphics[width=1.0\linewidth]{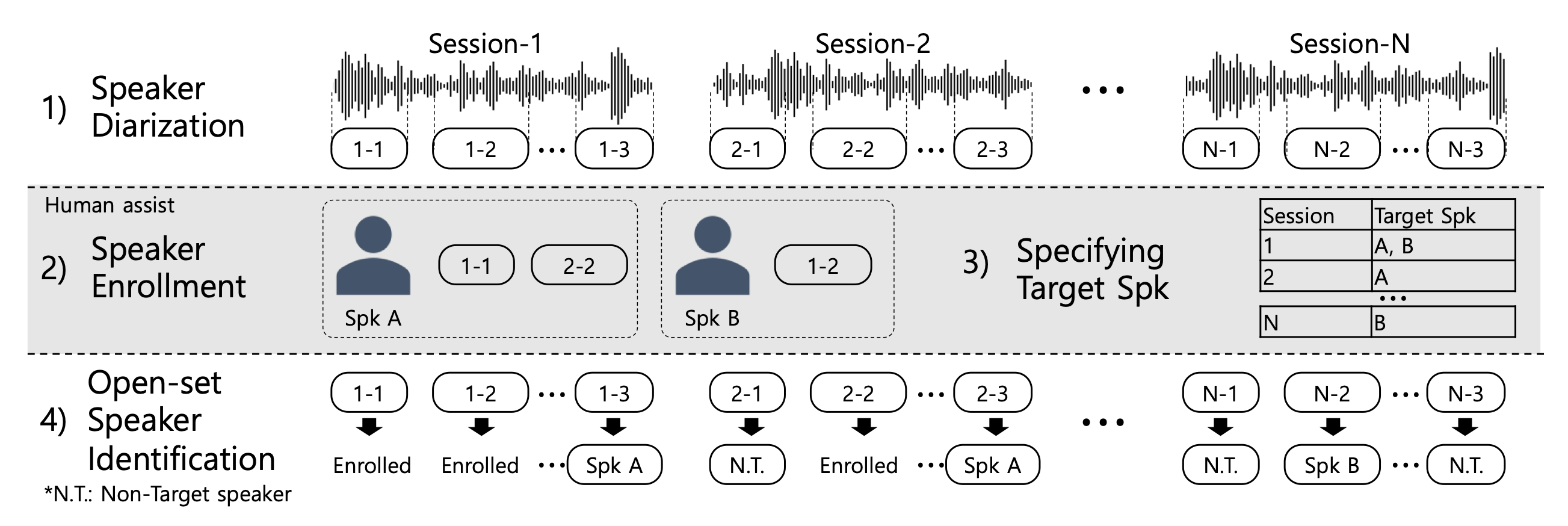}
    \caption{Overview of the target speaker tagging system: (1)~speaker diarization segments audio into single-speaker regions, (2)~a user enrolls known speakers, (3)~target speakers present in the session are specified, and (4)~the system assigns enrolled speaker identities to matching segments.}
    \label{fig:overall}
\end{figure*}

Figure~\ref{fig:overall} illustrates the full scenario: (1)~speaker diarization segments the audio into single-speaker regions, (2)~a user selects segments of their own speech for enrollment, (3)~the set of target speakers present in the session is specified, and (4)~the system assigns enrolled speaker identities to matching segments. For evaluation, we simulate the human enrollment step by selecting appropriate segments based on ground-truth annotations (details in Sections~\ref{sec:source} and~\ref{sec:datasets}), enabling reproducible assessment of system performance under realistic conditions.

\subsection{Speaker Diarization}
\label{sec:diarization}

Speaker diarization determines ``who spoke when'' in an audio recording~\cite{anguera2012speaker, park2022review}. In TST, its primary role is to split the audio into single-speaker segments that become the fundamental units for subsequent tagging. Most systems operate through three steps: (1)~detecting speech segments, (2)~extracting speaker embeddings, and (3)~clustering embeddings to group segments by speaker. Unlike standalone diarization, the accuracy of this step in TST strongly influences overall performance, bringing additional considerations.

\newpara{Clustering error.} Under-clustering and over-clustering are well-known problems~\cite{sinclair2013challenges, evans2012comparative}. Rather than minimizing diarization errors in isolation, we consider how each error type impacts downstream stages. Under-clustering, where different speakers are merged into the same cluster, causes critical problems for speaker identification. Over-clustering, where a single speaker is split into multiple clusters, can sometimes be corrected at the identification stage by merging segments assigned to the same enrolled speaker. Experimental results in Section~\ref{sec:main_results} confirm this asymmetry.

\newpara{Length of segments.} Longer segments generally produce higher-quality speaker embeddings, but in meeting scenarios, they also increase the risk of speaker changes within a single segment~\cite{poddar2018speaker, jung2019short}. Tuning an appropriate margin length for each segment is therefore necessary (Section~\ref{sec:main_results}).

\subsection{Open-Set Speaker Identification}
\label{sec:opensetsid}

Open-set speaker identification~\cite{malegaonkar2011performance, peri2023voxwatch, lin2024voxblink2} determines whether a given utterance belongs to one of the target speakers or not. In TST, its role is to assign an enrolled speaker ID to each diarized segment when possible.

The procedure is as follows: speaker embeddings are extracted from each segment and from the enrolled speech of each target speaker. The system calculates similarity scores between each segment embedding and target speaker embeddings. If the highest-scoring match exceeds a predefined threshold, the segment is labeled with the corresponding speaker ID; otherwise, it is tagged as non-target.

\newpara{Compensation on short segments.} Unlike classic open-set identification, TST can leverage other segments within the same session. By combining embeddings from multiple segments sharing the same diarization label and high similarity, the system approximates extracting a speaker embedding from a longer, more representative utterance, improving overall embedding quality. Results for this technique are presented in Section~\ref{sec:main_results}.

%==================================================================================
\section{TST-Bench: Synthetic Benchmark}
\label{sec:tstbench}

\subsection{Motivation}
\label{sec:motivation}

Table~\ref{tab:dataset_comparison} summarizes the properties of existing datasets in relation to TST requirements. While corpora such as ICSI~\cite{janin2003icsi, janin2004icsi} and AMI~\cite{carletta2005ami} provide global speaker identifiers, they contain a limited number of speakers in total, making it difficult to study how performance scales with the enrolled population. Challenge datasets such as DIHARD~\cite{ryant2019second} offer diverse acoustic conditions but use session-local labels without cross-session speaker identifiers. Standard speaker recognition benchmarks such as VoxCeleb~\cite{Nagrani17, chung2018voxceleb2, Nagrani19} contain many speakers but consist of short, pre-segmented utterances unsuitable for diarization-dependent evaluation. VoxBlink2~\cite{lin2024voxblink2} extends this line by providing large-scale open-set speaker identification evaluation, but similarly relies on pre-segmented utterances. Furthermore, the number of speakers per session in existing corpora is often limited or biased toward small group settings (e.g., 3--8 in ICSI, 4 in AMI). This is a critical limitation for TST evaluation: a larger number of speakers per session increases the candidate pool during identification, raising the probability that a non-target segment is erroneously matched to an enrolled speaker above the acceptance threshold. Evaluating whether a TST system can maintain accuracy under such conditions requires datasets with sufficiently large per-session speaker counts, which existing corpora lack. None of these resources support the full range of TST evaluation scenarios.

To address these limitations, we construct TST-Bench, a large-scale synthetic benchmark designed specifically for target speaker tagging evaluation. TST-Bench supports 8--30 speakers per session, enabling systematic study of how per-session speaker density affects diarization, identification, and non-target rejection. Synthetic data generation offers several advantages: (1)~perfect ground-truth annotations are available by construction, (2)~conditions such as number of speakers, and noise level can be precisely controlled, and (3)~data can be generated at scale without privacy concerns. We acknowledge several limitations of synthetic data. First, the source recordings consist of read speech from audiobooks, which differs from the spontaneous, conversational speaking style typical of real meetings. Second, the turn layout is generated algorithmically and does not replicate the complex dynamics of natural turn-taking, such as backchannels, interruptions, and floor-holding patterns. Third, mixing clean single-speaker recordings onto background noise does not fully reproduce the acoustic conditions of real meeting rooms, including reverberation, crosstalk, and far-field effects. Despite these gaps, TST-Bench provides an essential controlled testbed for systematic evaluation: as shown in Section~\ref{sec:main_results}, experiments on the ICSI Meeting Corpus confirm that the performance trends observed on synthetic data are consistent with those on real-world recordings.

\begin{table}[t]
\centering
\caption{Comparison of existing datasets with TST-Bench in terms of TST suitability. ``Global ID'' indicates cross-session speaker labels. ``Multi-Sess.'' indicates speakers appearing across multiple sessions.}
\label{tab:dataset_comparison}
\setlength{\tabcolsep}{3.5pt}
\begin{tabular}{lccccc}
\toprule
Dataset & Type & \# Spk & \makecell{Global\\ID} & \makecell{Multi-\\Sess.} & \makecell{TST\\Support} \\
\midrule
ICSI~\cite{janin2003icsi}     & Real  & 53    & \checkmark & \checkmark & Partial \\
AMI~\cite{carletta2005ami}     & Real  & 187   & \checkmark & Limited    & Partial \\
DIHARD~\cite{ryant2019second}  & Real  & --    & --         & --         & --      \\
VoxCeleb~\cite{Nagrani17}      & Real  & 1,251 & \checkmark & \checkmark & --      \\
\midrule
TST-Bench                       & Synth & 350   & \checkmark & \checkmark & \checkmark \\
\bottomrule
\end{tabular}
\end{table}

\subsection{Source Corpus and Speaker Inventory}
\label{sec:source}

TST-Bench is constructed from single-speaker recordings sourced from the Multilingual LibriSpeech (MLS) corpus~\cite{pratap2020mls}. A key design requirement for TST-Bench is that each speaker must contribute a sufficient amount of unique speech across multiple sessions, since speech segments are consumed without reuse during synthesis. To satisfy this requirement, we randomly select 350 English speakers from MLS, each having at least one hour of recorded speech.

Since MLS is a crowdsourced corpus, each speaker's recordings tend to be captured with a specific recording device and acoustic environment, causing speaker identity and channel characteristics to be confounded. Directly using these recordings for recognition risks discriminating speakers based on channel differences rather than genuine vocal characteristics. To mitigate this issue, we apply speech enhancement to all source recordings before synthesis using Resemble Enhance~\footnote{\url{https://github.com/resemble-ai/resemble-enhance}}, a publicly available denoising tool, thereby reducing device- and environment-dependent variation.

We then obtain word-level time boundaries using the Montreal Forced Aligner (MFA)~\cite{mcauliffe2017montreal}\footnote{\url{https://montreal-forced-aligner.readthedocs.io/}}. Speech segments are defined by grouping consecutive voiced intervals, with silence gaps exceeding 0.3~seconds treated as segment boundaries. This ensures that only clean, single-speaker speech intervals are used for synthesis.

The speaker pool is partitioned into two disjoint groups:
\begin{itemize}
\item \textbf{Enrolled speakers} ($N_e = 150$): Speakers whose identities are known to the system. Their speech appears in synthesized sessions alongside unknown speakers. In a real deployment scenario, a user would listen to a session and select segments containing their own speech for enrollment. Since ground-truth speaker labels are available by construction in TST-Bench, we simulate this process by automatically selecting segments belonging to each enrolled speaker from the synthesized sessions. For each enrolled speaker, approximately 20~seconds of speech-only audio is selected as enrollment data and excluded from the evaluation set to avoid overlap between enrollment and test conditions.
\item \textbf{Unknown speakers} ($N_u = 200$): Speakers not enrolled in the system. Their utterances appear in sessions but should be tagged as non-target.
\end{itemize}
The partitioning is performed randomly with the constraint that all selected enrolled speakers must have sufficient speech data to appear across multiple sessions and to provide at least 20~seconds of enrollment material within those sessions.

\subsection{Session Synthesis Pipeline}
\label{sec:synthesis}

\begin{figure}[t]
    \centering
    \includegraphics[width=\columnwidth]{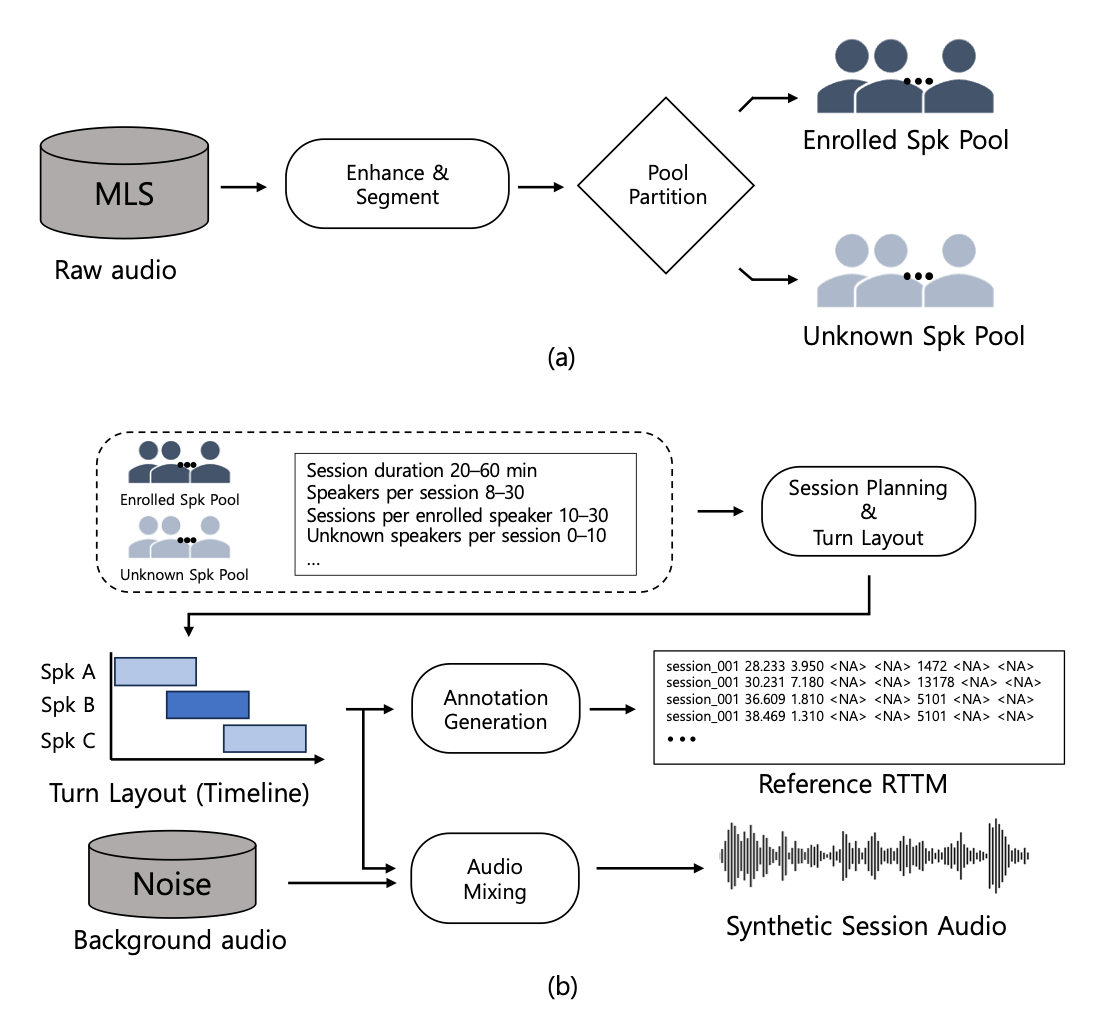}
    \caption{TST-Bench synthesis pipeline. (a)~Source data preparation: single-speaker recordings from MLS are enhanced, segmented via forced alignment, and partitioned into enrolled and unknown speaker pools. (b)~Session synthesis: speakers are sampled per session, turns are laid out with Dirichlet-distributed proportions, and mixed onto background noise to produce multi-speaker audio with global RTTM annotations.}
    \label{fig:synthesis}
\end{figure}

Each synthetic session simulates a multi-speaker meeting. The synthesis pipeline consists of four stages: session planning, turn layout, audio mixing, and annotation generation. Figure~\ref{fig:synthesis} illustrates the overall pipeline.

\newpara{Session planning.} For each session, the following parameters are sampled: session duration (uniformly between 20 and 60 minutes), the number of participating speakers (8--30 per session, including both enrolled and unknown speakers). Each session is guaranteed to contain at least one enrolled speaker. Each enrolled speaker is assigned to appear in 10--30 sessions to ensure sufficient representation, while 0--10 unknown speakers are included per session to create realistic non-target conditions.

\newpara{Turn layout.} Within each session, speaker turns are arranged sequentially with configurable inter-turn gaps (0.15--2.5~seconds). Turn durations range from 0.5 to 15 seconds. In real conversations, speaker participation is rarely uniform: some speakers dominate while others contribute infrequently. To reflect this, the proportion of total speaking time allocated to each speaker within a session is sampled from a symmetric Dirichlet distribution~\cite{kotz2019continuous} with concentration parameter $\alpha = 3.0$. This value produces moderate variability in speaker proportions---allowing dominant speakers to emerge naturally while avoiding extreme cases where a single speaker monopolizes the session or all speakers contribute equally.

\newpara{Audio mixing.} Speech segments are drawn from the source corpus without replacement, ensuring no speech content is reused across sessions. Rather than placing speech onto silence, each session uses a continuous background noise track onto which speech segments are mixed at a specified signal-to-noise ratio (SNR), sampled uniformly between 0 and 10~dB per session.

Since each session requires a single consistent background noise spanning 20--60 minutes, substantially longer noise recordings are needed than those typically available in standard noise corpora. To obtain such recordings, we source ambient noise from Freesound\footnote{\url{https://freesound.org/}}, downloading recordings of at least 3 minutes in duration using search queries such as ``meeting room silence,'' ``library ambience,'' and ``hallway noise,'' restricted to recordings with permissive licenses. Because the presence of speech in background noise would compromise the completeness of ground-truth annotations, we filter candidates using an in-house voice activity detection (VAD) engine and retain only recordings where detected speech constitutes less than 5\% of the total duration. When a selected noise recording is shorter than the session duration, it is repeated and concatenated; to avoid audible discontinuities at concatenation points, fade-in and fade-out effects are applied at each boundary. All audio is rendered at 16~kHz mono.

\newpara{Annotation generation.} For each session, an RTTM file is generated containing the precise onset, duration, and \textit{global} speaker label for every speech segment. Global labels ensure that the same speaker appearing in different sessions is consistently identified (e.g., speaker ``SPK0042'' retains the same label across all sessions in which they appear). Each label is additionally marked as either enrolled or unknown to facilitate scenario-specific evaluation.

\subsection{Dataset Statistics}
\label{sec:statistics}

Table~\ref{tab:tstbench_stats} summarizes the default TST-Bench configuration. The benchmark provides 300 sessions totaling approximately 200 hours of audio, with each session containing 8--30 speakers drawn from a pool of 350.\footnote{The pre-generated TST-Bench dataset and evaluation protocols are publicly available.}

\begin{table}[t]
\centering
\caption{TST-Bench default configuration and statistics.}
\label{tab:tstbench_stats}
\setlength{\tabcolsep}{4pt}
\begin{tabular}{ll}
\toprule
Parameter & Value \\
\midrule
Source corpus            & MLS (English)~\cite{pratap2020mls}    \\
Total speakers           & 350                   \\
Enrolled speakers ($N_e$)       & 150                      \\
Unknown speakers ($N_u$)        & 200                      \\
Enrollment duration      & 20~s / speaker             \\
Number of sessions       & 300                        \\
Session duration         & 20--60 min                 \\
Speakers per session     & 8--30                      \\
Sessions per enrolled speaker & 10--30                \\
Unknown speakers per session & 0--10                  \\
Background noise SNR     & 0--10 dB                   \\
Audio format             & 16~kHz, mono               \\
Annotation               & RTTM (global labels)       \\
\bottomrule
\end{tabular}
\end{table}

%==================================================================================
\section{Evaluation Protocol}
\label{sec:protocol}

\subsection{Metrics}
\label{sec:metrics}

We evaluate TST using the Detection and Identification Rate (DIR) and False Alarm Rate (FAR)~\cite{jain2011handbook}, defined using the notation in Table~\ref{tab:notations}. Let $\mathbf{D}$ denote the set of evaluated segments, partitioned into target segments $\mathbf{D}_T$ (from enrolled speakers) and non-target segments $\mathbf{D}_N$ (from unknown speakers):
\begin{equation}
\begin{aligned}
&\text{DIR}(\theta) = \frac{|\{d_t \mid \text{id}_{\text{tag}}(d_t, \theta) = \text{id}_{\text{ans}}(d_t),\, d_t \in \mathbf{D}_T\}|}{|\mathbf{D}_T|} \\
&\text{FAR}(\theta) = \frac{|\{d_n \mid \text{score}_{max}(d_n) \ge \theta,\, d_n \in \mathbf{D}_N\}|}{|\mathbf{D}_N|}
\end{aligned}
\label{eq:dir_far}
\end{equation}

DIR quantifies the ability to correctly detect and identify enrolled speakers: a target segment $d_t$ is considered correct only if the assigned identity matches the true identity and the similarity score exceeds threshold $\theta$. FAR measures the failure to reject unknown speakers: a non-target segment $d_n$ is a false alarm if its highest similarity score exceeds $\theta$. Since the decision threshold $\theta$ governs the DIR--FAR trade-off, we report DIR at fixed FAR operating points (DIR@FAR)~\cite{lin2024voxblink2}. For diarization quality, we additionally report Diarization Error Rate (DER)~\cite{anguera2012speaker}.

\begin{table}[t]
\centering
\caption{Notation used in the evaluation protocol.}
\label{tab:notations}
\setlength{\tabcolsep}{2pt}
\resizebox{\columnwidth}{!}{%
\begin{tabular}{ccl}
\toprule
$\mathbf{D}$  &   $\triangleq$  &   Set of evaluated single-speaker segments    \\
$\mathbf{D}_{T}$  &   $\triangleq$  &   $\mathbf{D}$ corresponding to target speakers   \\
$\mathbf{D}_{N}$  &   $\triangleq$  &   $\mathbf{D}$ corresponding to non-target speakers    \\
$\theta$  &   $\triangleq$  &   Predefined threshold  \\
$\text{score}_{max}(d)$   &   $\triangleq$  & Highest similarity score of $d$ on gallery \\
$\text{id}_{\text{ans}}(d)$&   $\triangleq$  & True identity of $d$   \\
$\widetilde{\text{id}_{\text{tag}}}(d)$ &   $\triangleq$  & Identity corresponding to $\text{score}_{max}(d)$ \\
$\text{id}_{\text{tag}}(d, \theta)$ &   $\triangleq$  & Tagged identity of $d$ \\
    & = & $\begin{cases}
\widetilde{\text{id}_{\text{tag}}}(d), & \text{if } \text{score}_{max}(d) \ge \theta \\
\text{non-target}, & \text{otherwise} \end{cases}$ \\
\bottomrule
\end{tabular}}
\end{table}

\subsection{Evaluation Scenarios}
\label{sec:scenarios}

TST-Bench defines two evaluation scenarios of increasing complexity: Scenario~1 evaluates speaker diarization in isolation, while Scenario~2 evaluates the full TST pipeline including speaker identification.

\newpara{Scenario 1: Speaker diarization.} This scenario performs standard DER evaluation on the synthesized sessions, isolating the segmentation component and enabling analysis of diarization quality independently of identification.

\newpara{Scenario 2: Full pipeline (TST).} The complete TST pipeline is executed: diarization produces segments, which are then identified against a \textit{session-specific gallery} of speakers present in the session. Since different diarization systems produce different segmentations, a common set of \textit{evaluation segments} must be defined to enable fair comparison.

\newpara{Evaluation segment selection.} Evaluation segments are derived from the reference RTTM by selecting non-overlapping single-speaker regions whose duration is at least $\tau_{\min}$ seconds. In this work, we set $\tau_{\min} = 1$~s. This threshold can be adjusted depending on the application; for instance, applications that do not require robust performance on very brief interjections or frequent turn changes may adopt a higher threshold. By fixing the evaluation segments based on the reference RTTM, the number and composition of evaluated segments remain constant regardless of the diarization system used.

Since system-generated segments do not necessarily align with these evaluation segments, naively evaluating at the granularity of system segments would cause the number of evaluated units to vary across diarization systems, making results incomparable. To address this, we propose a reference-anchored evaluation protocol. Each evaluation segment (defined from the reference RTTM) is matched to the system segment with which it has the longest temporal overlap, and the identified speaker label of the matched system segment is transferred to the evaluation segment.

Two notable cases arise under this mapping. First, a single long system segment may be matched to multiple evaluation segments. Because each system segment carries a single speaker label, only the evaluation segments whose ground-truth identity matches that label are counted as correct; the rest are penalized. This naturally reflects the cost of under-clustering, where merging distinct speakers into one segment causes misidentification for all but one of the true speakers. Second, an evaluation segment with no overlapping system segment---typically due to missed speech detection---receives no identification result. Since it remains in the denominator of DIR but cannot contribute to the numerator, such misses directly lower DIR, ensuring that diarization failures propagate appropriately into the overall TST metric.

This protocol ensures that the set of evaluated segments and their ground-truth conditions remain fixed regardless of the diarization system used, enabling fair comparison across different systems.

%==================================================================================
\section{Experimental Setup}
\label{sec:setup}

\subsection{Datasets}
\label{sec:datasets}

We evaluate on two datasets. For real data, we use the \textbf{ICSI Meeting Corpus}~\cite{janin2003icsi, janin2004icsi}, processed following the protocol described below. For synthetic data, we use \textbf{TST-Bench} with the default configuration (Table~\ref{tab:tstbench_stats}).

In ICSI, we select speakers appearing in 2--10 sessions as known speakers. Enrollment segments are drawn from up to two sessions per speaker, selecting the ten longest utterances per session. All sessions containing at least one known speaker serve as test sessions. From these, we extract evaluation segments that are (i)~not used for enrollment, (ii)~free of overlapping speech, and (iii)~at least 1 second in duration, consistent with the evaluation segment selection protocol described in Section~\ref{sec:scenarios}. Table~\ref{tab:data} compares the statistics of the resulting evaluation sets.

\begin{table}[t]
\centering
\caption{Statistics of evaluation datasets. VoxCeleb is included for reference as a widely-used speaker verification benchmark.}
\label{tab:data}
\setlength{\tabcolsep}{4pt}
\begin{tabular}{lcccc}
\toprule
                       & ICSI$^*$  & TST-Bench & Vox1-O & Vox1-H  \\
\midrule
\# of utterances       & 52,141 & 204,042  & 4,708   & 137,924 \\
Avg dur. of utt. (s) & 3.07  & 2.80    & 8.28   & 8.25    \\
Min dur. of utt. (s) & 1.00  & 1.00    & 3.96  & 3.96     \\
Max dur. of utt. (s) & 100.00+ & 22.68   & 69.04  & 100.00+  \\
\bottomrule
\multicolumn{5}{l}{\small $^*$Processed evaluation set derived from ICSI.}
\end{tabular}
\end{table}

\subsection{Implementation Details}
\label{sec:implementation}

The speaker diarization component is built on a high-resolution embedding extractor (HEE)~\cite{heo2023high} trained on VoxCeleb1 and VoxCeleb2~\cite{Nagrani17, chung2018voxceleb2, Nagrani19}. Embeddings are enhanced with dimensionality reduction and attention-based aggregation, followed by spectral clustering~\cite{heo2020clova, kwon2021adapting}.

For open-set speaker identification, we employ an ECAPA-TDNN model~\cite{desplanques2020ecapa, das2021hlt, chung2020defence} also trained on VoxCeleb1 and VoxCeleb2. We use a 4-second window and a 1.5-second shift to extract 256-dimensional speaker embeddings. When a segment or enrollment utterance yields multiple embeddings, score-level average pooling is applied. Similarity scores are computed using cosine similarity followed by adaptive symmetric normalization (AS-Norm)~\cite{cumani2011comparison}. The cohort set for score normalization consists of utterances from 2,000 speakers randomly selected from VoxBlink2~\cite{lin2024voxblink2}, with an adaptation size of 20.

%==================================================================================
\section{Experimental Results}
\label{sec:results}

\subsection{Main Results}
\label{sec:main_results}

We present results on both TST-Bench and the ICSI corpus, analyzing the impact of different system components and comparing trends across synthetic and real-world data.

\newpara{Evaluation across scenarios.} We evaluate the two scenarios defined in Section~\ref{sec:scenarios}. For Scenario~1 (speaker diarization), the baseline system achieves a DER of 8.70\% on TST-Bench, decomposed into 6.23\% miss, 0.35\% false alarm, and 2.12\% speaker confusion, with a Jaccard Error Rate (JER) of 15.66\%. The dominant error source is missed speech (6.23\%), primarily caused by overlapping segments in the synthesized sessions. Since the baseline diarization system does not perform explicit overlap detection, temporally overlapping speech from multiple speakers is attributed to only one speaker, inflating the miss rate.

For Scenario~2 (full pipeline), the baseline system achieves DIR@FAR of 88.79\% at FAR=0.5\%, 93.00\% at FAR=1\%, 96.80\% at FAR=5\%, and 97.61\% at FAR=10\%. These results serve as the starting point for the ablation studies presented below. Note that the baseline system with no segmentation margin corresponds to a naive integration of speaker diarization and open-set speaker identification without TST-specific adaptations.

\newpara{Effect of clustering tendency.} While lower diarization error might intuitively suggest better downstream performance, the type of clustering error matters more than its magnitude for TST. To demonstrate this, we re-run Scenario~2 under three diarization configurations that differ in clustering tendency. We measure speaker confusion alongside two clustering quality metrics: homogeneity and completeness~\cite{rosenberg2007v}. Homogeneity quantifies whether each predicted cluster contains only segments from a single speaker; low homogeneity indicates under-clustering, where distinct speakers are merged into the same cluster. Completeness quantifies whether all segments from a given speaker are assigned to the same cluster; low completeness indicates over-clustering, where a single speaker is split across multiple clusters.

Table~\ref{tab:clustering_tendency} presents the results. The baseline configuration balances homogeneity and completeness (0.851 and 0.850, respectively) and achieves the lowest speaker confusion (2.12\%). Deliberately inducing under-clustering lowers homogeneity (0.834) while maintaining completeness (0.854), and increases confusion to 3.04\%; this degrades DIR@FAR=0.5\% to 86.75\%. Conversely, inducing over-clustering raises homogeneity (0.854) at the cost of completeness (0.836) with a comparable confusion increase (3.30\%), yet DIR@FAR=0.5\% improves to 89.46\%, surpassing even the baseline. This confirms that over-clustering is preferable for TST: split clusters can be re-merged at the identification stage, whereas under-clustering irreversibly contaminates segment embeddings with mixed-speaker information. The ICSI results in Table~\ref{tab:clustering_tendency} corroborate this trend: over-clustering achieves the highest DIR@FAR=0.5\% (94.63\%), followed by under-clustering (94.57\%) and baseline (94.51\%). While the absolute differences are smaller than on TST-Bench, the relative ordering is consistent. The narrower gaps on ICSI can be attributed to the smaller number of speakers per session (3--8 vs.\ 8--30 in TST-Bench), which reduces the likelihood and severity of under-clustering errors.

\begin{table}[t]
\centering
\caption{Effect of clustering tendency on TST-Bench and ICSI (Scenario~2). Homo.\ and Comp.\ denote homogeneity and completeness, respectively.}
\label{tab:clustering_tendency}
\setlength{\tabcolsep}{4pt}
\resizebox{\columnwidth}{!}{%
\begin{tabular}{llcccc}
\toprule
Dataset & Configuration & \makecell{Spk.\\Conf.\ [\%]} & Homo. & Comp. & \makecell{DIR@\\FAR=0.5\%} \\
\midrule
\multirow{3}{*}{TST-Bench} & Baseline        & 2.12 & 0.851 & 0.850 & 88.79 \\
 & Under-cluster & \textbf{3.04} & 0.834 & 0.854 & 86.75 \\
 & Over-cluster & 3.30 & 0.854 & 0.836 & \textbf{89.46} \\
\midrule
\multirow{3}{*}{ICSI} & Baseline & 1.81 & 0.625 & 0.627 & 94.51 \\
 & Under-cluster & \textbf{1.31} & 0.621 & 0.643 & 94.57 \\
 & Over-cluster & 3.33 & 0.631 & 0.594 & \textbf{94.63} \\
\bottomrule
\end{tabular}}
\end{table}

\newpara{Effect of segment margins.} Diarization systems typically detect speech boundaries tightly to minimize false alarms in speech activity detection. We hypothesize that such tight boundaries may discard useful acoustic context, degrading the quality of speaker embeddings extracted near segment edges. To test this, we add symmetric margins of up to a specified maximum duration (e.g., 0.5~s) to the start and end of each segment. To prevent overlap between adjacent segments, the margin is adjusted on a per-segment basis so that no two segments overlap after expansion. Table~\ref{tab:tstbench_margin} presents the results.

The results show that on TST-Bench, a small margin of 0.1~s improves DIR@FAR=0.5\% from 88.79\% to 89.05\%, confirming that slightly extended segments produce more robust speaker embeddings. Larger margins (0.25~s and 0.5~s) yield marginal or no further improvement, suggesting that a modest expansion suffices to capture sufficient acoustic context around speech boundaries. The ICSI results follow the same trend at strict operating points: margins progressively improve DIR@FAR=0.5\% from 94.51\% (no margin) to 94.86\% (0.25~s). Interestingly, at lenient operating points (FAR=5\% and 10\%), margins slightly degrade performance on ICSI (e.g., 98.88\% to 98.14\% at FAR=5\%), likely because excessive expansion risks capturing neighboring speakers' audio in the tightly packed turn-taking of real meetings.

\begin{table}[t]
\centering
\caption{Effect of segment margins on TST-Bench and ICSI (Scenario~2). Margins are added symmetrically and adjusted per segment to avoid overlap.}
\label{tab:tstbench_margin}
\setlength{\tabcolsep}{5pt}
\begin{tabular}{llcccc}
\toprule
\multirow{2}{*}{Dataset} & \multirow{2}{*}{Margin} & \multicolumn{4}{c}{DIR@FAR [\%]} \\ \cmidrule(l){3-6}
 & & 0.5\% & 1\% & 5\% & 10\% \\
\midrule
\multirow{4}{*}{TST-Bench} & None   & 88.79 & 93.00 & 96.80 & 97.61 \\
 & 0.1~s  & \textbf{89.05} & 92.98 & 96.78 & \textbf{97.63} \\
 & 0.25~s & 88.88 & \textbf{93.04} & 96.81 & 97.62 \\
 & 0.5~s  & 88.71 & 93.02 & \textbf{96.83} & 97.62 \\
\midrule
\multirow{4}{*}{ICSI} & None & 94.51 & 94.67 & \textbf{98.88} & \textbf{99.26} \\
 & 0.1~s  & 94.69 & 94.82 & 98.70 & 99.01 \\
 & 0.25~s & \textbf{94.86} & 94.99 & 98.43 & 98.79 \\
 & 0.5~s  & 94.83 & \textbf{95.14} & 98.14 & 98.55 \\
\bottomrule
\end{tabular}
\end{table}

\newpara{Short-utterance compensation.} In conventional speaker recognition, evaluation is performed on pre-segmented utterances of sufficient length. In TST, however, speaker identification operates on diarization output, which inevitably includes segments of varying lengths. Short segments tend to produce lower-quality speaker embeddings, degrading identification accuracy. One advantage of the TST pipeline is that diarization labels provide grouping information: segments sharing the same label are presumed to belong to the same speaker. By selecting the top-$N$ most similar segments within the same diarization label and combining their embeddings, the system can compensate for short-segment degradation.

Table~\ref{tab:tstbench_comp} presents the results. On TST-Bench, the top-$N$ merging strategy progressively improves DIR@FAR across all operating points, with top-3 achieving 89.03\% at FAR=0.5\%, compared to 88.79\% without compensation. We also evaluate a label-based strategy that assigns a single speaker ID to all segments sharing the same diarization label. While this approach achieves the highest DIR at lenient operating points (e.g., 97.40\% at FAR=5\%), it suffers a sharp performance drop at strict thresholds (81.82\% at FAR=0.5\%), as diarization errors propagate directly to the tagging output. This indicates that while leveraging diarization labels is beneficial, full label-level aggregation amplifies the impact of clustering mistakes, particularly under-clustering errors where segments from different speakers share the same label.

The ICSI results confirm the same progressive improvement trend with top-$N$ merging. Notably, however, the label-based strategy does not exhibit the dramatic performance collapse observed on TST-Bench (94.85\% vs.\ 81.82\% at FAR=0.5\%). This is likely because ICSI sessions contain fewer speakers, making severe under-clustering---the primary failure mode of label-based aggregation---less prevalent. Another notable observation on ICSI is that DIR@FAR=0.5\% and DIR@FAR=1\% are nearly identical across all compensation methods (e.g., 94.82\% at both operating points for top-3). This plateau arises because ICSI's smaller non-target population yields a sparse score distribution, so widening the FAR threshold from 0.5\% to 1\% does not change which segments exceed the acceptance boundary, effectively producing the same DIR. TST-Bench, with its larger speaker pool and more non-target segments, provides a denser score distribution that differentiates operating points more clearly (e.g., 89.03\% vs.\ 94.15\% for top-3 at FAR=0.5\% and 1\%, respectively), underscoring the value of a large-scale benchmark for fine-grained evaluation.

\begin{table}[t]
\centering
\caption{Short-utterance compensation on TST-Bench and ICSI (Scenario~2). Label-based assigns a single ID to all segments sharing a diarization label.}
\label{tab:tstbench_comp}
\setlength{\tabcolsep}{5pt}
\resizebox{\columnwidth}{!}{%
\begin{tabular}{llcccc}
\toprule
\multirow{2}{*}{Dataset} & \multirow{2}{*}{Method} & \multicolumn{4}{c}{DIR@FAR [\%]} \\ \cmidrule(l){3-6}
 & & 0.5\% & 1\% & 5\% & 10\% \\
\midrule
\multirow{5}{*}{TST-Bench} & No compensation & 88.79 & 93.00 & 96.80 & 97.61 \\
 & Top-1           & 88.95 & 93.39 & 96.98 & 97.72 \\
 & Top-2           & 88.94 & 93.78 & 97.10 & 97.79 \\
 & Top-3           & \textbf{89.03} & \textbf{94.15} & 97.21 & \textbf{97.85} \\
 & Label-based     & 81.82 & 95.32 & \textbf{97.40} & 97.78 \\
\midrule
\multirow{5}{*}{ICSI} & No compensation & 94.51 & 94.67 & 98.88 & 99.26 \\
 & Top-1           & 94.73 & 94.73 & 98.97 & 99.24 \\
 & Top-2           & 94.80 & 94.80 & 98.99 & 99.26 \\
 & Top-3           & 94.82 & 94.82 & 99.04 & 99.27 \\
 & Label-based     & \textbf{94.85} & \textbf{94.85} & \textbf{99.21} & \textbf{99.30} \\
\bottomrule
\end{tabular}}
\end{table}

\subsection{Discussion}
\label{sec:discussion}

The experimental results on both TST-Bench and ICSI yield consistent insights. TST performance is not merely a function of individual component quality; the interaction between diarization and identification creates compound effects not visible when evaluating each in isolation. The clustering experiments on both TST-Bench and ICSI confirm that optimizing diarization metrics alone does not guarantee optimal TST performance, and that dedicated design choices---such as preferring over-clustering and leveraging short-utterance compensation---yield meaningful improvements.

Compared to evaluating on existing corpora, TST-Bench enables controlled experiments that disentangle these factors. Its controlled conditions allow researchers to systematically study each dimension of difficulty, facilitating targeted improvements. While synthetic data cannot fully replace real-world evaluation, it provides a necessary complement for scalable and reproducible benchmarking.

%==================================================================================
\section{Conclusion}
\label{sec:conclusion}

We have presented target speaker tagging (TST), a task that integrates speaker diarization, verification, and identification into a unified framework for multi-speaker conversations. Through experiments on both TST-Bench and the ICSI Meeting Corpus, we demonstrated that TST requires dedicated system design: diarization metrics do not always align with tagging objectives, and techniques such as over-clustering preference and short-utterance compensation yield meaningful improvements.

To address the critical absence of evaluation resources for TST, we introduced TST-Bench, a large-scale synthetic benchmark with over 150 enrolled speakers, 300 sessions, and configurable conditions. Our comprehensive evaluation protocol, spanning diarization and full-pipeline scenarios, reveals that TST poses challenges not captured by conventional benchmarks. The interplay between diarization quality, identification accuracy, and unknown speaker rejection creates a multi-dimensional difficulty landscape that existing metrics fail to characterize.

We publicly release the TST-Bench dataset and evaluation scripts to foster reproducible research on this important and under-explored task. We hope this resource encourages the community to move beyond isolated task-specific benchmarks and toward integrated evaluation frameworks that better reflect the complexity of real-world speaker recognition applications.

\section{Generative AI Use Disclosure}

A generative AI assistant was used to refine grammar and improve the clarity of expressions in the manuscript originally drafted by the authors. All technical content, experimental design, and scientific claims were produced entirely by the authors.

%==================================================================================
% References begin here (allowed to start before page 9 if space permits)
\bibliographystyle{IEEEtran}
\bibliography{mybib}

@string{icassp = {Proc. {IEEE} Int. Conf. Acoust., Speech, Signal Process. ({ICASSP})}}

@string{is = {Proc. {INTERSPEECH}}}

@book{kotz2019continuous,
  title={Continuous multivariate distributions, Volume 1: Models and applications},
  author={Kotz, Samuel and Balakrishnan, Narayanaswamy and Johnson, Norman L},
  volume={1},
  year={2019},
  publisher={John wiley \& sons}
}

@inproceedings{mcauliffe2017montreal,
  title={Montreal Forced Aligner: Trainable Text-Speech Alignment Using Kaldi},
  author={McAuliffe, Michael and Socolof, Michaela and Mihuc, Sarah and Wagner, Michael and Sonderegger, Morgan},
  booktitle=is,
  pages={498--502},
  year={2017}
}

@inproceedings{rosenberg2007v,
  title={V-measure: A conditional entropy-based external cluster evaluation measure},
  author={Rosenberg, Andrew and Hirschberg, Julia},
  booktitle={Proceedings of the 2007 joint conference on empirical methods in natural language processing and computational natural language learning (EMNLP-CoNLL)},
  pages={410--420},
  year={2007}
}

@inproceedings{cumani2011comparison,
  title={Comparison of speaker recognition approaches for real applications},
  author={Cumani, Sandro and Batzu, Pier Domenico and Colibro, Daniele and Vair, Claudio and Laface, Pietro and Vasilakakis, Vasileios},
  booktitle=is,
  pages={2365--2368},
  year={2011}
}

@inproceedings{snyder2018x,
  title={X-vectors: Robust dnn embeddings for speaker recognition},
  author={Snyder, David and Garcia-Romero, Daniel and Sell, Gregory and Povey, Daniel and Khudanpur, Sanjeev},
  booktitle=icassp,
  pages={5329--5333},
  year={2018}
}

@inproceedings{carletta2005ami,
  title={The AMI meeting corpus: A pre-announcement},
  author={Carletta, Jean and Ashby, Simone and Bourban, Sebastien and Flynn, Mike and Guillemot, Mael and Hain, Thomas and Kadlec, Jaroslav and Karaiskos, Vasilis and Kraaij, Wessel and Kronenthal, Melissa and others},
  booktitle={International workshop on machine learning for multimodal interaction},
  pages={28--39},
  year={2005},
  organization={Springer}
}

@inproceedings{ryant2019second,
  title={The Second DIHARD Diarization Challenge: Dataset, Task, and Baselines},
  author={Ryant, Neville and Church, Kenneth and Cieri, Christopher and Cristia, Alejandrina and Du, Jun and Ganapathy, Sriram and Liberman, Mark},
  booktitle=is,
  pages={978--982},
  year={2019}
}

@inproceedings{pratap2020mls,
  title={MLS: A Large-Scale Multilingual Dataset for Speech Research},
  author={Pratap, Vineel and Xu, Qiantong and Sriram, Anuroop and Synnaeve, Gabriel and Collobert, Ronan},
  booktitle=is,
  pages={2757--2761},
  year={2020}
}

@inproceedings{lin2024voxblink2,
  title={VoxBlink2: A 100K+ Speaker Recognition Corpus and the Open-Set Speaker-Identification Benchmark},
  author={Lin, Yuke and Cheng, Ming and Zhang, Fulin and Gao, Yingying and Zhang, Shilei and Li, Ming},
  booktitle=is,
  pages={4263--4267},
  year={2024}
}

@book{jain2011handbook,
  title={Handbook of face recognition},
  author={Jain, Anil K and Li, Stan Z},
  volume={1},
  number={2},
  year={2011},
  publisher={Springer}
}

@inproceedings{janin2003icsi,
  title={The ICSI meeting corpus},
  author={Janin, Adam and Baron, Don and Edwards, Jane and Ellis, Dan and Gelbart, David and Morgan, Nelson and Peskin, Barbara and Pfau, Thilo and Shriberg, Elizabeth and Stolcke, Andreas and others},
  booktitle=icassp,
  volume={1},
  pages={I--I},
  year={2003}
}

@article{anguera2012speaker,
  title={Speaker diarization: A review of recent research},
  author={Anguera, Xavier and Bozonnet, Simon and Evans, Nicholas and Fredouille, Corinne and Friedland, Gerald and Vinyals, Oriol},
  journal={IEEE Transactions on audio, speech, and language processing},
  volume={20},
  number={2},
  pages={356--370},
  year={2012},
  publisher={IEEE}
}

@article{park2022review,
  title={A review of speaker diarization: Recent advances with deep learning},
  author={Park, Tae Jin and Kanda, Naoyuki and Dimitriadis, Dimitrios and Han, Kyu J and Watanabe, Shinji and Narayanan, Shrikanth},
  journal={Computer Speech \& Language},
  volume={72},
  pages={101317},
  year={2022},
  publisher={Elsevier}
}

@article{peri2023voxwatch,
  title={VoxWatch: an open-set speaker recognition benchmark on VoxCeleb},
  author={Peri, Raghuveer and Sadjadi, Seyed Omid and Garcia-Romero, Daniel},
  journal={arXiv preprint arXiv:2307.00169},
  year={2023}
}

@inproceedings{malegaonkar2011performance,
  title={Performance evaluation in open-set speaker identification},
  author={Malegaonkar, Amit and Ariyaeeinia, Aladdin},
  booktitle={Biometrics and ID Management: COST 2101 European Workshop, BioID 2011, Brandenburg (Havel), Germany, March 8-10, 2011. Proceedings 3},
  pages={106--112},
  year={2011},
  organization={Springer}
}

@article{campbell2002speaker,
  title={Speaker recognition: A tutorial},
  author={Campbell, Joseph P},
  journal={Proceedings of the IEEE},
  volume={85},
  number={9},
  pages={1437--1462},
  year={2002},
  publisher={IEEE}
}

@article{bai2021speaker,
  title={Speaker recognition based on deep learning: An overview},
  author={Bai, Zhongxin and Zhang, Xiao-Lei},
  journal={Neural Networks},
  volume={140},
  pages={65--99},
  year={2021},
  publisher={Elsevier}
}

@inproceedings{heo2023high,
  title={High-resolution embedding extractor for speaker diarisation},
  author={Heo, Hee-Soo and Kwon, Youngki and Lee, Bong-Jin and Kim, You Jin and Jung, Jee-weon},
  booktitle=icassp,
  pages={1--5},
  year={2023}
}

@inproceedings{variani2014deep,
  title={Deep neural networks for small footprint text-dependent speaker verification},
  author={Variani, Ehsan and Lei, Xin and McDermott, Erik and Moreno, Ignacio Lopez and Gonzalez-Dominguez, Javier},
  booktitle=icassp,
  pages={4052--4056},
  year={2014}
}

@article{dehak2010front,
  title={Front-end factor analysis for speaker verification},
  author={Dehak, Najim and Kenny, Patrick J and Dehak, R{\'e}da and Dumouchel, Pierre and Ouellet, Pierre},
  journal={IEEE Transactions on Audio, Speech, and Language Processing},
  volume={19},
  number={4},
  pages={788--798},
  year={2010},
  publisher={IEEE}
}

@article{heo2020clova,
  title={Clova baseline system for the voxceleb speaker recognition challenge 2020},
  author={Heo, Hee Soo and Lee, Bong-Jin and Huh, Jaesung and Chung, Joon Son},
  journal={arXiv preprint arXiv:2009.14153},
  year={2020}
}

@Article{Nagrani19,
              author       = "Arsha Nagrani and Joon~Son Chung and Weidi Xie and Andrew Zisserman",
              title        = "Voxceleb: Large-scale speaker verification in the wild",
              journal      = "Computer Science and Language",
              year         = "2019",
              publisher    = "Elsevier",
            }

@InProceedings{chung2018voxceleb2,
  title={VoxCeleb2: Deep Speaker Recognition},
  author={Chung, Joon Son and Nagrani, Arsha and Zisserman, Andrew},
  booktitle=is,
  pages={1086--1090},
  year={2018}
}

@InProceedings{Nagrani17,
	author       = "Nagrani, A. and Chung, J.~S. and Zisserman, A.",
	title        = "VoxCeleb: a large-scale speaker identification dataset",
	booktitle    = is,
	year         = "2017",
}

@inproceedings{desplanques2020ecapa,
  title={ECAPA-TDNN: Emphasized Channel Attention, Propagation and Aggregation in TDNN Based Speaker Verification},
  author={Desplanques, Brecht and Thienpondt, Jenthe and Demuynck, Kris},
  booktitle=is,
  pages={1--5},
  year={2020}
}

@inproceedings{chung2020defence,
  title={In Defence of Metric Learning for Speaker Recognition},
  author={Chung, Joon Son and Huh, Jaesung and Mun, Seongkyu and Lee, Minjae and Heo, Hee-Soo and Choe, Soyeon and Ham, Chiheon and Jung, Sunghwan and Lee, Bong-Jin and Han, Icksang},
  booktitle=is,
  pages={2977--2981},
  year={2020}
}

@inproceedings{fujita2019end,
  title={End-to-end neural speaker diarization with self-attention},
  author={Fujita, Yusuke and Kanda, Naoyuki and Horiguchi, Shota and Xue, Yawen and Nagamatsu, Kenji and Watanabe, Shinji},
  booktitle={2019 IEEE Automatic Speech Recognition and Understanding Workshop (ASRU)},
  pages={296--303},
  year={2019},
  organization={IEEE}
}

@inproceedings{sinclair2013challenges,
  title={Where are the challenges in speaker diarization?},
  author={Sinclair, Mark and King, Simon},
  booktitle=icassp,
  pages={7741--7745},
  year={2013}
}

@article{evans2012comparative,
  title={A comparative study of bottom-up and top-down approaches to speaker diarization},
  author={Evans, Nicholas and Bozonnet, Simon and Wang, Dong and Fredouille, Corinne and Troncy, Rapha{\"e}l},
  journal={IEEE Transactions on Audio, speech, and language processing},
  volume={20},
  number={2},
  pages={382--392},
  year={2012},
  publisher={IEEE}
}

@article{poddar2018speaker,
  title={Speaker verification with short utterances: a review of challenges, trends and opportunities},
  author={Poddar, Arnab and Sahidullah, Md and Saha, Goutam},
  journal={IET Biometrics},
  volume={7},
  number={2},
  pages={91--101},
  year={2018},
  publisher={Wiley Online Library}
}

@inproceedings{jung2019short,
  title={Short utterance compensation in speaker verification via cosine-based teacher-student learning of speaker embeddings},
  author={Jung, Jee-weon and Heo, Hee-Soo and Shim, Hye-jin and Yu, Ha-Jin},
  booktitle={2019 IEEE automatic speech recognition and understanding workshop (ASRU)},
  pages={335--341},
  year={2019},
  organization={IEEE}
}

@inproceedings{janin2004icsi,
  title={The ICSI meeting project: Resources and research},
  author={Janin, Adam and Ang, Jeremy and Bhagat, Sonali and Dhillon, Rajdip and Edwards, Jane and Macias-Guarasa, Javier and Morgan, Nelson and Peskin, Barbara and Shriberg, Elizabeth and Stolcke, Andreas and others},
  booktitle={Proceedings of the 2004 ICASSP NIST Meeting Recognition Workshop},
  year={2004}
}

@inproceedings{kwon2021adapting,
  title={Adapting speaker embeddings for speaker diarisation},
  author={Kwon, Youngki and Jung, Jee Weon and Heo, Hee Soo and Kim, You Jin and Lee, Bong Jin and Chung, Joon Son},
  booktitle=is,
  pages={2493--2497},
  year={2021}
}

@article{das2021hlt,
  title={HLT-NUS SUBMISSION FOR 2020 NIST Conversational Telephone Speech SRE},
  author={Das, Rohan Kumar and Tao, Ruijie and Li, Haizhou},
  journal={arXiv preprint arXiv:2111.06671},
  year={2021}
}

@inproceedings{rouvier2015speaker,
  title={Speaker diarization through speaker embeddings},
  author={Rouvier, Mickael and Bousquet, Pierre-Michel and Favre, Benoit},
  booktitle={2015 23rd european signal processing conference (eusipco)},
  pages={2082--2086},
  year={2015},
  organization={IEEE}
}

@article{wang2021dku,
  title={The DKU-Duke-Lenovo system description for the third DIHARD speech diarization challenge},
  author={Wang, Weiqing and Lin, Qingjian and Cai, Danwei and Yang, Lin and Li, Ming},
  journal={arXiv preprint arXiv:2102.03649},
  year={2021}
}

@article{park2019auto,
  title={Auto-tuning spectral clustering for speaker diarization using normalized maximum eigengap},
  author={Park, Tae Jin and Han, Kyu J and Kumar, Manoj and Narayanan, Shrikanth},
  journal={IEEE Signal Processing Letters},
  volume={27},
  pages={381--385},
  year={2019},
  publisher={IEEE}
}

\end{document}